\documentclass[useAMS,usenatbib]{mn2e}
\usepackage{graphicx,amsmath}
\usepackage{natbib}

\title[Evolution and chemical yields of AGB stars:
effects of low-temperature opacities]{Evolution and chemical yields of AGB stars:
effects of low-temperature opacities}
\author[P. Ventura and P. Marigo]{P. Ventura$^{1}$\thanks{E-mail:
ventura@oa-roma.inaf.it (AVR)} and P. Marigo$^{2}$\\
$^{1}$INAF-Osservatorio Astronomico di Roma, Via Frascati 33, Monte Porzio Catone 00040, Italia\\
$^{2}$ Department of Astronomy, University of Padova,
        Vicolo dell'Osservatorio 3, I-35122 Padova, Italy}
\begin{document}

\date{Accepted 1988 December 15. Received 1988 December 14; in original form 1988 October 11}

\pagerange{\pageref{firstpage}--\pageref{lastpage}} \pubyear{2002}

\maketitle

\label{firstpage}

\begin{abstract}
The studies focused on the Thermally-Pulsing Asymptotic Giant Branch
phase experienced by low- and intermediate-mass stars are
extremely important in many astrophysical contexts. In particular, 
a detailed computation of their chemical yields 
is essential for several issues, ranging from the chemical evolution of galaxies, 
to the mechanisms behind the formation of globular clusters. 
Among all the uncertainties affecting the theoretical modelling of this
phase, and described in the literature, it remains to be fully clarified 
which results are severely affected by the use of inadequate low-temperature opacities, 
that are in most cases calculated on the basis 
of the original chemical composition of the stars, and do not consider the changes 
in the surface chemistry due to the occurrence of the third dredge-up and hot-bottom burning. 
Our investigation is aimed at investigating  this point.
By means of full evolutionary models including   
new set of molecular opacities computed specifically with the \AE SOPUS tool, 
we highlight which stellar models, among 
those present in the literature, need a substantial revision, mainly 
in relation to the predicted chemical yields.  The interplay among convection,
hot bottom burning and the low-temperature opacity treatment is also discussed.
\end{abstract}


\section{Introduction}
After the end of helium burning in the core, 
stars with initial masses $M$ in the range $1\la M/M_{\odot} \la 6-8$ 
evolve through the Thermally-Pulsing Asymptotic Giant Branch (hereinafter TP-AGB) phase during 
which they experience a very rich nucleosynthesis whose
products (e.g. He, C, N, Ne, Mg, s-process elements) 
may be convected to outermost layers and  eventually 
ejected into the interstellar medium (ISM) by stellar winds, 
leaving a CO white dwarf remnant and an expanding planetary nebula.
All these facts make the study of AGB stars 
extremely important for the chemical evolution of the host system.

In the last decades, the interest towards AGB evolution has increased essentially
for two reasons: a) The observations of high-redshift systems allowed
the measurement of carbon and nitrogen abundances; b) AGB stars have been
suggested as the main responsible for the self-enrichment of globular
clusters, to explain the existence of multiple populations detected in 
almost all the clusters examined in the Galaxy on the basis of the
photometric and spectroscopic surveys \citep{norris,paolo3}. 

Chemical yields of AGB stars critically depend by the  
variations of their surface chemical composition, that may be produced not
only by the first (and possibly the second) dredge-up, but also by two
further mechanisms that are typical of the TP-AGB phase: a) The Hot Bottom
Burning (hereinafter HBB), i.e. the occurrence of proton capture
reactions at the bottom of the external convective mantle, as a consequence
of the high temperatures achieved by these layers in the most massive models;
b) the Third Dredge-up (hereinafter TDU), by which the bottom of the envelope
after each TP sinks inwards into layers previously touched by $3\alpha$ 
nucleosynthesis. Both mechanisms alter the surface chemical composition: 
as for HBB we  expect abundance changes resulting from the CNO-cycle  
reactions operating at equilibrium, 
whereas the TDU is associated to a net increase of the surface carbon and, 
to a smaller extent, of oxygen.

The predictive power of the yields provided by the theoretical investigations
is unfortunately undermined by the many uncertainties in the input-physics
that affect the AGB description: mass loss, convection, treatment of the
convective/radiative interfaces are unknown from first principles, and all
have a strong impact on the physical (and consequently chemical) evolution
of AGB models \citep{falk}. These problems are the reason for the great 
differences among the results found by different research groups on this 
topic \citep{paola01,fenner,amanda,paolo4}.

A further uncertainty, still to be fully investigated in this context, is
the treatment of the molecular opacities, that are commonly calculated on the
basis of the assumed metal content  and the initial abundance distribution 
of the elements, 
neglecting any alteration of the surface chemistry associated to HBB and/or
TDU. 

An exploratory approach related to this problem was made by \citet{paola02},
who showed that when the C/O ratio exceeds unity, the formation of CN 
molecules determines an increase of the opacity, that is expected to favour 
an expansion of the envelope, an increase of mass loss, and a faster
consumption of the whole external mantle: this would diminish the number 
of TPs experienced by the star, and the effects of the TDU would be 
considerably reduced.
 
In a  follow-up paper \cite{paola07} considered the possible effects driven 
by variations of molecular
opacities in the most massive AGB stars
(with $M \ga 3.5  M_{\odot}$),  which experience both TDU and HBB.
With the aid of envelope integrations it was found that if the dredge-up of carbon 
is efficient enough to lead to an early transition from C/O $\,<1$ to C/O $\,>1$, then
hot-bottom burning may be weakened, extinguished, or even prevented.
 
A step forward in understanding the effects 
produced by the use of opacity tables appropriate to carbon-rich
mixtures was made by \citet{sergio}, who presented a study focused on
a $2M_{\odot}$ model of metallicity $Z=10^{-4}$. The authors concluded that the surface 
composition and the global yields by low-mass AGB stars of low metallicity 
are significantly affected by the use of the C-rich opacities.
More extended sets of full TP-AGB evolutionary calculations with variable molecular
opacities have been recently presented by  Weiss \& Ferguson (2009), covering wide ranges 
of stellar masses and metallicities.

In the present paper we  further extend the investigation 
on the effects of low-temperature opacities on the AGB evolution,
so as to single out the regimes of stellar masses for which a major revision of 
the published models is required.

\section[]{Description of the Models}
\subsection{The evolution code}
The models presented in this paper were calculated by means of the ATON code
for stellar evolution, a full description of which can be found in \citet{paolo1}. 
We provide here a brief summary of the numerical and physical 
inputs used in the present computations.
Convection was modelled according to the Full Spectrum of Turbulence (hereinafter
FST) prescription by \citet{canuto}. Mixing of chemicals and nuclear burning were 
coupled by means of a diffusive approach, following the scheme presented
in \citet{clout}; accordingly, extra-mixing was modelled by an exponential decay of
convective velocities beyond the formal borders, with an e-folding decay
of $l=\zeta H_P$. During the two main phases of core nuclear burning and 
in occurrence of the second dredge-up we used $\zeta=0.02$, whereas in
the TPs phase no extra-mixing was considered. Mass loss was modelled according
to \citet{blo}, with the free parameter entering Reimers' prescription $\eta_R=0.02$,
in agreement with the calibration given in \citet{paolo2}.

\subsection{Molecular opacities}
A key implementation in the present stellar models resides in the use
of new tables of low-temperature opacities ($1500 \le T \le 30\, 000$ K), which are
suitably constructed to follow the changes in the chemical composition of the
envelope driven by TDU and HBB. 

A large set of tables of Rosseland mean (RM) opacities has been computed with a new tool, 
\AE SOPUS (Accurate Equation of State and OPacity Utility Software) 
described in Marigo \& Aringer (2009; 
web-interface at http://stev.oapd.inaf.it/aesopus), 
to which the reader should refer to all the
details.
Each table of RM opacities covers a 
rectangular grid in the $\log(T)$-$\log(R)$ 
diagram, where the temperature is made vary in the interval $3.2 \le \log(T)
\le 4.5$, while the $R= \rho \,(10^6/T[{\rm K}])^3$ parameter spans the range $-8 \le \log(R) \le
1$. Following the formalism introduced by Marigo \& Aringer (2009), the adopted 
reference mixture 
assumes a metallicity 
$Z_{\rm ref}=0.001$,  and an $\alpha$-enhanced distribution of the
elements, 
expressed by the parameter $[\alpha/{\rm Fe}]=+0.4$, with respect to the 
reference solar mixture given by the compilation of \citet{gs98}.

According to the notation in Marigo \& Aringer (2009), the assumed 
$\alpha$-enhanced composition corresponds to mixture $A$, 
in which the abundances of $\alpha$-elements  (O, Ne, Mg, Si, S, Ca, and Ti)
are enhanced while the metallicity is kept constant, $Z=Z_{\rm ref}$.
The latter condition is fulfilled by 
requiring that the abundance variation 
of the enhanced elements is compensated by the total
abundance variation of all the others.

This implies that the concentrations of O and all other
$\alpha$-elements are incremented by $\sim 23\%$ compared to the scaled-solar
values, while the abundances of the iron-group elements and
all other metals, including C and N, are depressed by almost a factor of $2$.
As a consequence, the reference chemical mixture -- with 
$Z_{\rm ref}=0.001$, $[\alpha/{\rm Fe}]=+0.4$ -- corresponds to
a carbon-to-oxygen ratio C/O$\sim 0.19$, which is quite lower than the 
(C/O)$_{\odot} \sim 0.49$ of the \cite{gs98} solar mixture.

As a next step, the reference $\alpha$-enhanced mixture is
further altered by varying the abundances of C, N and O, in order to
account for the changes of these elements caused by TDU and HBB. 
To this aim we introduce three independent variation factors $f_i$ for
 C/O,  C, and N defined as:
\begin{eqnarray*}
\left(\displaystyle\frac{X_{\rm C}}{X_{\rm O}}\right)  & = & 
f_{\rm C/O} \, \left(\displaystyle\frac{X_ {\rm C, ref}}{X_{\rm O, ref}}\right) \\
X_{\rm C} & = &  f_{\rm C}\, X_{\rm C, ref} \\
X_{\rm N} & = &  f_{\rm N}\, X_{\rm N, ref}\, , 
\end{eqnarray*}
where 
($X_{\rm C, ref}$, $X_{\rm N, ref}$, $X_{\rm O,
ref}$) correspond to the abundances (in mass fraction) of carbon, nitrogen and oxygen in the
reference chemical mixture with $Z=Z_{\rm ref}=0.001$ and $[\alpha/{\rm Fe}]=+0.4$, while  
($X_{\rm C}$, $X_{\rm N}$, $X_{\rm O}$)
denote the abundances of the same elements in the new mixture.
By construction, the variation factor of
oxygen derives from  condition: 
\begin{equation*}
X_{\rm O}  =  f_{\rm O}\, X_{\rm O, \rm ref} \,\,\,\,\,\,{\rm
  with}\,\,\,\,\,\,f_{\rm O}  =  f_{\rm C} / f_{\rm C/O}\\
\end{equation*}
We notice that in this case the reference metallicity is not preserved, and the total metal
content $Z$ may be larger or lower than $Z_{\rm ref}$ depending on the total CNO abundance variation.
\begin{figure}
\resizebox{1.\hsize}{!}{\includegraphics{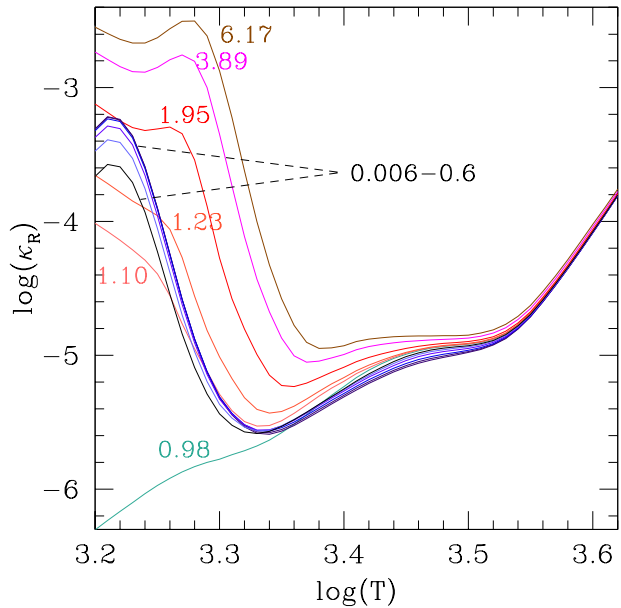}}
\caption{Rosseland mean opacity as a function of the temperature for several C/O values, as 
indicated. We assume $\log(R)=-5$. The abundance of carbon is made vary, while leaving unchanged that of oxygen.
The reference chemical mixture is defined 
by $Z_{\rm ref}=0.001$, $X=0.7$, $[\alpha/{\rm Fe}]=+0.4$,
and solar abundances from Grevesse \& Sauval (1998).}
\label{fig_opac_co1}
\end{figure}

The last composition parameter we need to specify is the hydrogen abundance $X$,
and the normalization condition is fulfilled by  requiring that the helium 
abundance is given by $Y=1-X-Z$.
The assumed values of the four independent composition parameters $X$, $f_{\rm C/O}$,
$f_{\rm C}$ and $f_{\rm O}$ are listed in Table~\ref{tab_opacpar}.
In total, $1296$ opacity tables have been computed for all combinations
of these  parameters.
For its relevance to the opacity results, the last column of Table~\ref{tab_opacpar} shows also the 
C/O ratio corresponding to each $f_{\rm C/O}$.

In Fig.~\ref{fig_opac_co1} each curve shows the predicted RM opacities
for a given value of the C/O ratio, spanning over three orders of magnitude, passing from 
$\sim 0.006$ to  $\sim 6$. While for $\log(T)>3.55$ the different  curves almost overlap 
as the Rosseland mean is controlled by the continuum opacity of hydrogen 
(bound-free and free-free transitions), for $\log(T)<3.5$ they start to deviate significantly.
As discussed in Marigo \& Aringer (2009), for $3.55 \la \log(T) \la 3.45$
the RM opacity is mainly affected by the anion H$^{-}$, Thomson electron scattering,
and CN molecular bands, while for $3.45 \la \log(T) \la 3.2$ the most important
absorbers are the oxygen-bearing molecules such as H$_2$O, VO, TiO for C/O$~<1$, 
while C-bearing molecules such as CN, C$_2$, C$_3$, HCN, and C$_2$H$_2$ control the opacity for
C/O$~>1$. 

The results presented in Fig.~\ref{fig_opac_co1} provide a case overview of the 
changes in opacities driven by changes in C/O consequent to TDU and HBB.
Assuming C/O$_{\rm ref}=0.19$ as the reference value,
the opacity curves with C/O $<\,0.19$ correspond to an effective decrease of the carbon abundance 
($\log(f_c)<0$), hence
representing the cases in which HBB burns carbon in favor of nitrogen; on the other hand 
the opacity  curves with C/O $>~0.19$ correspond to an effective increase 
of the carbon abundance ($\log(f_c)>0$), hence suitable to describe the cases in which TDU prevails. 

One striking feature concerns the different sensitiveness of the RM opacities to changes in carbon
abundance. The diversification among the opacity curves is notably larger in mixtures with C/O$>1$,
while it keeps rather small as long as  C/O$\la 0.9$, at least  for $\log(T)\sim 3.4-3.6$.
As discussed in Sect.~\ref{sec_results}, this feature  has important consequences 
for the behaviour of the TP-AGB models, i.e.  C-rich models are in general more importantly 
affected by changes in the chemistry than O-rich models.
Finally, we remark that the adopted set of $f_{\rm C/O}$ and $f_{\rm C}$
accounts for both a decrease and an increase of C/O and C (hence also O).
Our choice has a well-founded motivation. In fact, during the TP-AGB evolution 
newly-synthesized C and O may be not only convected to the surface by TDU, but
their envelope abundances may be also depleted by HBB, i.e. partially burnt in favour of N by the
CNO-cycle. In this case the evolution of the surface C/O ratio does not simply trace the increase
of C due to TDU, being rather the result of concurring and competing effects (TDU and HBB), 
which should be considered in RM opacities suitable for AGB modelling.
\begin{table}
\caption{Values of the composition parameters $f_i$ adopted in the
computation of the RM opacity tables. The sequence of the variation
factor $f_{\rm C/O}$ is designed to allow a good sampling of the
critical points of C/O at which significant opacity changes are
expected. Refer to the text for more explanation.}
\label{tab_opacpar}
\begin{tabular}{crrrl}
\hline
$X$ &  $\log(f_{\rm C/O})$ & $\log(f_{\rm C})$ & $\log(f_{\rm N})$ & C/O \\
\hline
$0.5$ & $-1.50$ & $-1.5$ & $0.0$ & $6.166\,10^{-3}$ \\
$0.7$ & $-1.00$ & $-0.5$ & $0.5$ & $1.950\,10^{-2}$ \\
$0.8$ & $-0.50$ &  $0.0$ & $1.0$ & $6.166\,10^{-2}$ \\
    &  $0.00$ &  $0.5$ & $1.5$ & $1.950\,10^{-1}$ \\
    &  $0.30$ &  $1.0$ & $2.0$ & $3.890\,10^{-1}$ \\
    &  $0.50$ &  $1.5$ & $2.5$ & $6.166\,10^{-1}$ \\
    &  $0.70$ &      &     & $9.772\,10^{-1}$\\ 
    &  $0.75$ &      &     & $1.096$ \\ 
    &  $0.80$ &      &     & $1.230$ \\ 
    &  $1.00$ &      &     & $1.950$ \\ 
    &  $1.30$ &      &     & $3.890$ \\ 
    &  $1.50$ &      &     & $6.166$ \\ 
\hline
\end{tabular}
\end{table}
\begin{center}
\begin{table*}
  \caption{Chemical yields and evolutionary properties of AGB stars}
  \begin{tabular}{ccccccccc}
 \hline
 \AE SOPUS low-T opacities & $M/M_{\odot}$ & $[^{12}$C/Fe] & $[^{14}$N/Fe] & $[^{16}$O/Fe] & C/O
     & R(CNO) & $T_{\rm bce}~({\rm MK})$ & $\log(R/R_{\odot})$ \\
 \hline
 Standard: reference mixture & 2.5 & 1.80 & 0.52 & 0.98 & 2.45 & 11.2 & 31 & 2.55 \\
          & 3.0 & 0.80 & 1.94 & 0.68 & 0.49 &  5.5 & 75 & 2.60 \\
          & 3.5 & 0.03 & 1.86 & 0.42 & 0.15 &  3.6 & 87 & 2.68 \\
 \hline
  New: with CNO variations & 2.5 & 1.68 & 0.52 & 0.98 & 1.86 &  9.3 & 18 & 2.73 \\
           & 3.0 & 1.22 & 1.33 & 0.68 & 1.29 &  4.5 & 62 & 2.75 \\
           & 3.5 & 0.14 & 1.87 & 0.47 & 0.17 &  3.8 & 85 & 2.70 \\

\hline
\end{tabular}
\end{table*}
\end{center}

\begin{figure*}
\begin{minipage}{0.33\textwidth}
\resizebox{1.\hsize}{!}{\includegraphics{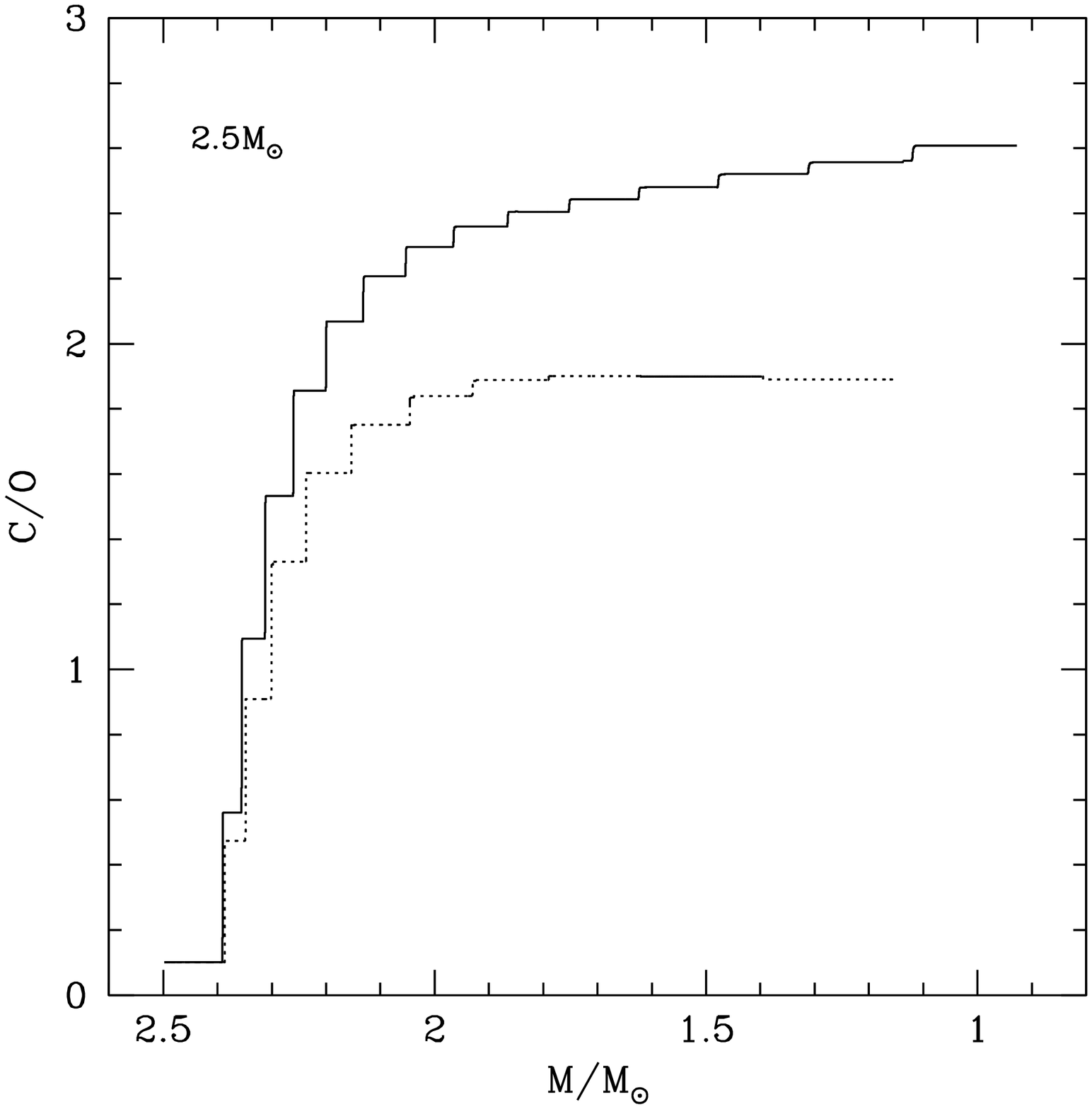}}
\end{minipage}
\begin{minipage}{0.33\textwidth}
\resizebox{1.\hsize}{!}{\includegraphics{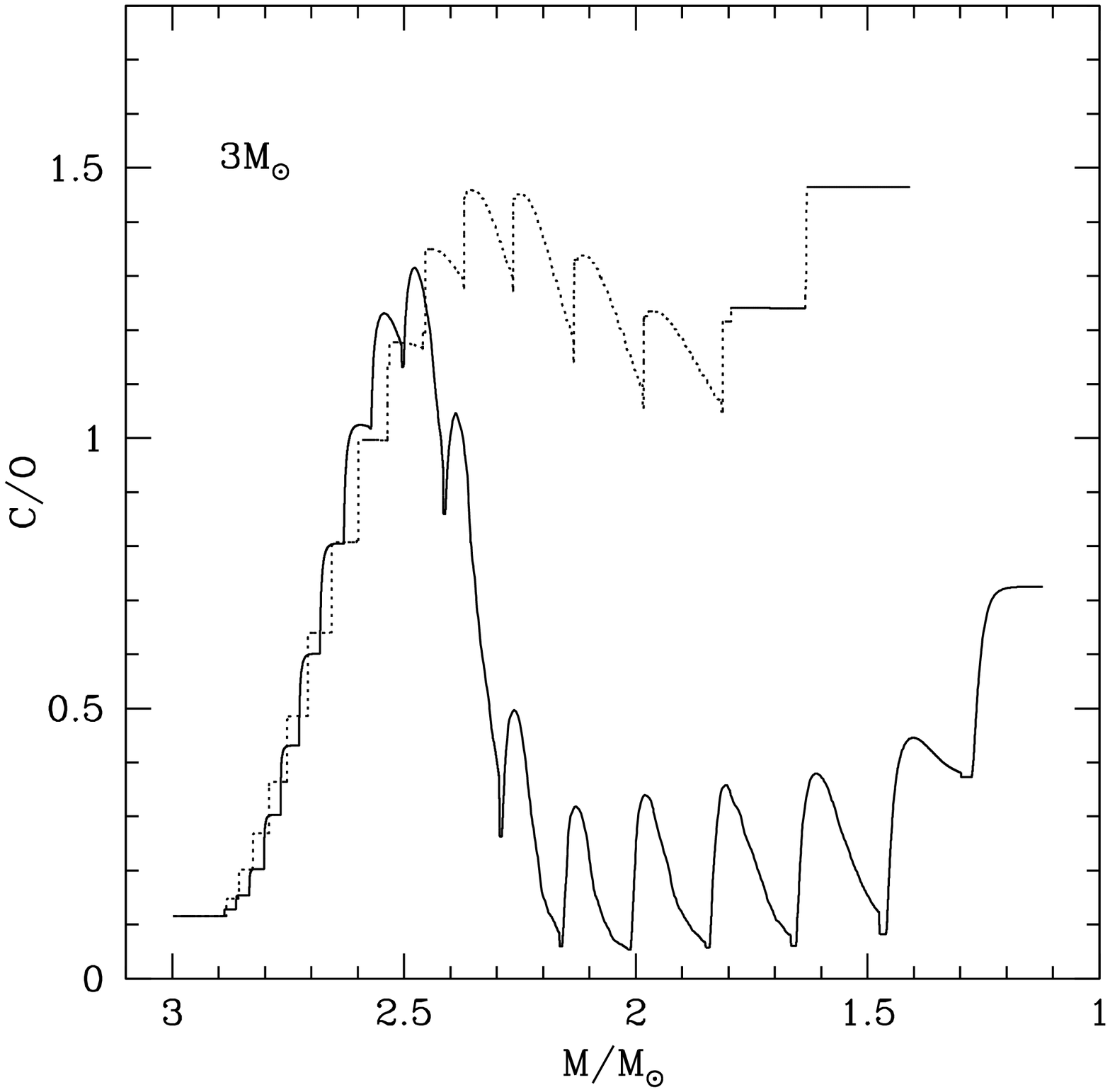}}
\end{minipage}
\begin{minipage}{0.33\textwidth}
\resizebox{1.\hsize}{!}{\includegraphics{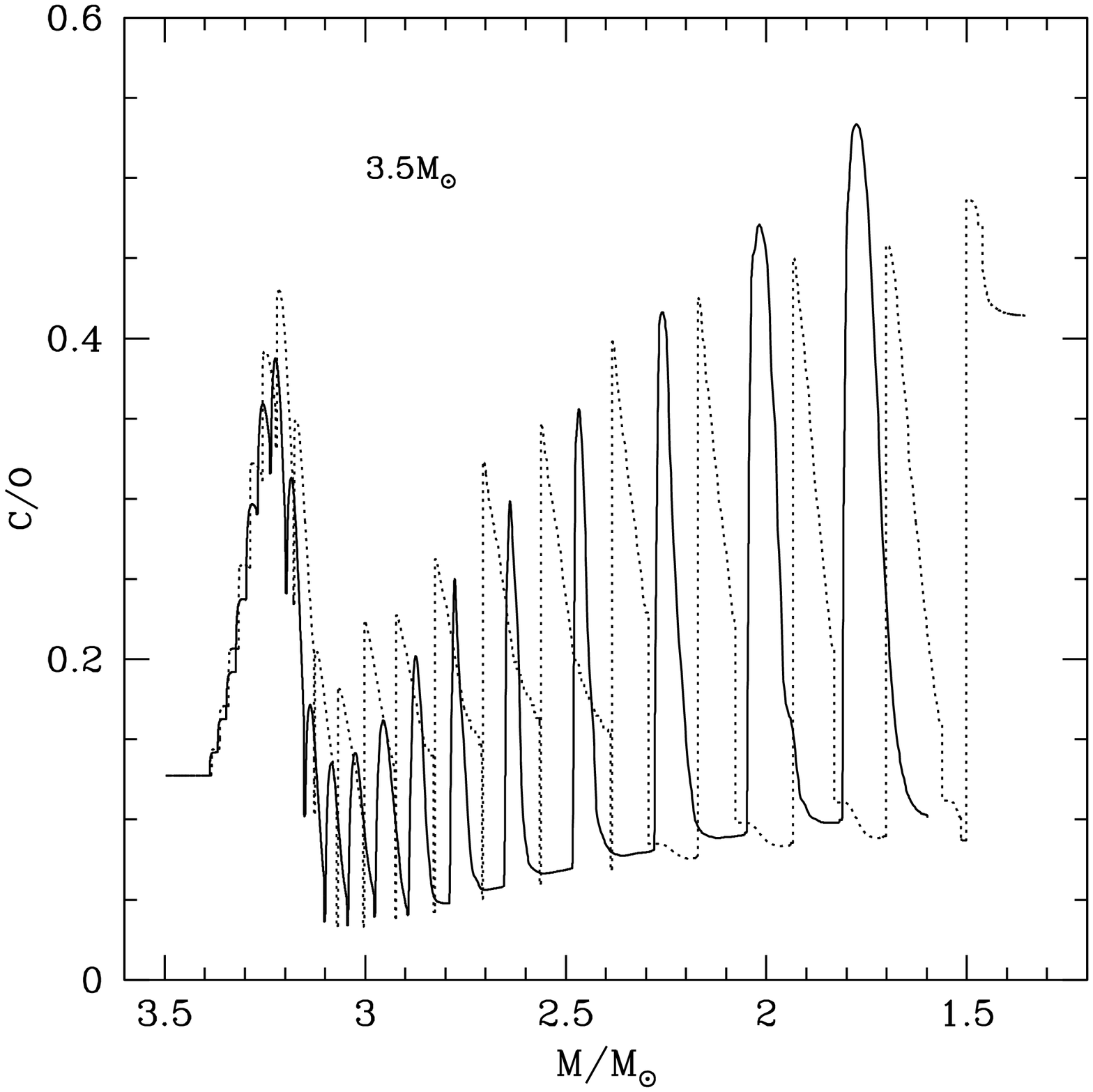}}
\end{minipage}
\caption{The evolution of the surface C/O ratio in three stellar
models calculated with the standard opacities (solid line), and 
with the new opacities that account for the changes in the CNO 
abundances (dotted line).}
\label{figco}
\end{figure*}
\section[]{The r\^{o}le of the new opacities}
\label{sec_results}
The evolutionary sequences presented in this work have been followed from the pre-MS
through the whole TP-AGB phase, up to the latest stages of the AGB evolution,
when almost all the envelope was lost. The initial chemical composition of the models is 
typical of an intermediate metallicity globular cluster: Z=0.001 and Y=0.24, 
with an alpha-enhanced ([$\alpha/$Fe]=0.4) distribution of the elements; 
the individual abundances of the various species are taken from \citet{gs98}. 
We focus our attention on three models of mass $M=2.5,\, 3.0,\, 3.5\,M_{\odot}$, as
their evolution encompass the whole range of cases that are relevant for
the present investigation. Each evolutionary model 
was calculated with the \AE SOPUS opacities assuming both the reference chemical 
mixture, hence neglecting the surface chemistry changes during the TP-AGB phase 
(``standard case''), and accounting for the CNO abundance variations 
(``new case'').

The chemical content of the ejecta, averaged over the whole AGB phase, 
are indicated in Table 2; R(CNO) in col.6 is the ratio between the overall 
CNO content of the ejecta compared to the initial value. Table 2 also includes, 
in the last two columns, the maximum temperature reached by the bottom of 
the convective envelope $T_{\rm bce}$ and the radius of the star 
before the envelope is lost.
We limit our discussion to the CNO elements here, because this allows
to better disentangle the effects of the opacity description from those
associated to the cross sections of the proton-capture reactions by heavier 
nuclei \citep{bob}.

Figure~\ref{figco} shows the evolution of the surface C/O ratio of the models.
The $2.5M_{\odot}$ model evolves as a carbon star. The use of the \AE SOPUS
CNO opacities favours larger radii (see last column of Table 1), with
a consequent increase of the mass loss rate: the faster consumption of the 
envelope prevents the C/O from exceeding 2, at
odds with the model calculated with the standard opacities, whose C/O ratio
approaches 3 in the final stages of the evolution. The average C/O of the
ejecta is consequently $\sim 40\%$ higher in the standard case. The cooling
effect of the CNO opacities is confirmed by the differences in the
maximum temperature attained by the bottom of the envelope; note that
these temperatures are well below the threshold ($\sim 60-70$K) necessary
to activate HBB, and this is the reason why the nitrogen content of
the ejecta is the same in the two cases.

The $3.0 M_{\odot}$ shows up the most striking differences (see the middle panel 
of fig.\ref{figco}). In the standard case we see an initial increase of the 
surface C/O, followed by a phase of carbon depletion, that is the signature of HBB.
The evolution of C/O in the \AE SOPUS CNO model is completely 
different: the increase of the surface carbon stops when C/O$\sim 1.5$, 
and it keeps approximately constant for the remaining 
evolution, the small depletion during the CNO burning phase being compensated 
by the increase determined by the TDU episodes following each TP. 
In this case the cooling of the envelope (see the different temperatures of 
the two models in the last column of Table 1) prevents HBB, so that this latter 
is extinguished. Unlike the lowest masses case, the yields 
of the standard model have a lower C/O ratio; also the nitrogen content of 
the ejecta is different, because HBB favours a larger increase of the nitrogen
abundance in the standard case. \\
The right panel of fig.\ref{figco} shows that in the $M=3.5M_{\odot}$ model
the differences introduced by the CNO opacities are small. This
is motivated by the low C/O ratio, never exceeding 0.5. The decrease of C/O 
after a few TPs indicates that HBB is active in both models. These results 
suggest that HBB, when strong, is efficient in both cases.
This fact is explained considering that in the range 
$0.05 \la {\rm C/O} \la 0.5$,  spanned by the $M=3.5M_{\odot}$ TP-AGB model, 
the predicted RM opacities differ relatively little, as displayed by the 
curves in Fig.~\ref{fig_opac_co1}.
\section{The demand for a revision of the AGB modelling}
The results presented in the previous section show that using the appropriate
low temperature opacities, that accounts for the formation of new molecules when 
the C/O ratio approaches and exceeds unity, may change substantially the
evolution of the surface abundances of carbon, nitrogen and oxygen during the
AGB evolution.
In low-mass AGB stars, when HBB is absent, the use of fixed opacities 
largely overestimate the carbon yield and the C/O ratio. 
On the other hand, we find  that for the models 
achieving mild HBB conditions, the treatment of the opacities is 
relevant to maintain or quench HBB itself, so that, unlike the previous case,
we expect much higher yields of $^{12}$C, and smaller $^{14}$N abundances. 
This effects is restricted 
to a limited range of masses, $\sim 0.5M_{\odot}$, because the results
for more massive models indicate that when a stronger HBB is present a
depletion of the surface carbon is achieved independently of the
opacity treatment; these models show some differences in the oxygen yields
(that is depleted more efficiently in the standard case), but even this
latter difference is expected to vanish with increasing mass.
These results
are in full agreement 
with the findings by Marigo (2007), obtained by means of synthetic TP-AGB calculations
based on numerical integrations of complete envelope models, 
which included an approximate description of variable molecular opacities (Marigo 2002).

The question of which TP-AGB models, among the most massive ones,
require substantial revision, as a consequence of introducing adequate molecular opacities,
is therefore shifted to the well-known and long-standing problems associated to the
AGB evolution: which conditions are necessary to achieve HBB \citep{paolo4}. 
It has been shown that HBB is strongly connected with the treatment of convection,
and that when this latter is modelled efficiently (e.g. when the FST model
is adopted) HBB is favoured \citep{blo2}. 
The results of the previous section indicate
that the range of masses whose yields are crucially affected by the new opacities
differs  depending on the convective modelling: when this latter is efficient,
only models for masses smaller that $\sim 3-3.5M_{\odot}$ the use of the
appropriate opacities is mandatory not to substantially mistake the estimation of the yields,
whereas AGB models computed with a low-efficiency convective model would demand even a
larger revision, extended to almost all the intermediate masses.
\section{Conclusions}
In this paper we focused on the r\^{o}le played by the treatment 
of low-temperature opacities, consistently coupled with the surface chemical composition, 
on the description of the AGB evolution. The scope was to clarify 
what part 
of the existing literature, and commonly used in many astrophysical contexts, needs
a severe revision due to the frequent incorrect estimate of the surface opacities.
To this scope, we present new evolutionary sequences at low metallicity
($Z=0.001$), that cover most
of the situations typical of the AGB phase in terms of TDU and HBB,
and compare the standard results with the new models, calculated
with a set of opacities appositely built for the present investigation, 
that accounts for the variation of the surface chemistry. 
Our main findings are the following:
\begin{enumerate}
\item In absence of HBB, the use of correct opacities is
mandatory, otherwise the carbon yields and the C/O ratio of the
ejecta are largely overestimated. This holds for all the low masses, 
that achieve TDU but not HBB.
\item For a narrow range of masses, those achieving only mild HBB
conditions, the necessity of using the new opacities is even more
urgent, because the surface evolution depends dramatically on the
opacity treatment: in this case HBB is quenched, carbon is never
depleted if not in minor quantities during the interpulse phase,
so that the traditional computations strongly underestimates the
carbon yields.
\item For more massive models, strong HBB prevents the surface C/O 
ratio to increase substantially, so that the opacity treatment seems 
 not critical for the results obtained.
\end{enumerate}
The general validity of these conclusions should be further tested by
extending the calculations at both higher and lower metallicities. 
This work is in progress. 
In any case,  thanks to the present availability of RM opacity tables
for arbitrary chemical mixtures,
(i.e. Marigo \& Aringer 2009 (the \AE SOPUS web-tool), 
Helling \& Lucas 2009, Lederer \& Aringer 2009), 
the use of molecular opacities, consistently connected to 
the actual surface chemistry, could become the standard choice 
for stellar evolutionary calculations to come.  
\section*{Acknowledgments}
P.M. acknowledges partial support from by the University of Padova
(60A02-2949/09), INAF/PRIN07 (CRA 1.06.10.03), and MIUR/PRIN07 (prot. 20075TP5K9).

\end{document}